\newcommand{\rem}[1]{}
\documentclass[prd,twocolumn,showpacs,showkeys,showpacs,preprintnumbers,nofootinbib,amsmath,amssymb]{revtex4-2}
\usepackage{amsfonts,amssymb,amsmath,mathrsfs}
\usepackage{url}
\usepackage{graphicx}
\newtheorem{thrm}{Theorem}

\newtheorem{prop}[thrm]{Proposition}

\newtheorem{remark}[thrm]{Remark}
\begin{document}
\title[Conserved currents in static CKG]{Note on conserved currents\\ in static Conformal Killing Gravity }
\author{Carlo Alberto Mantica}
\email{carlo.mantica@mi.infn.it} 
\author{Luca Guido Molinari} 
\email{luca.molinari@mi.infn.it}
\affiliation{Physics Department Aldo Pontremoli,
Universit\`a degli Studi di Milano and I.N.F.N. sezione di Milano,
Via Celoria 16, 20133 Milano, Italy.}
\date{\today}

\begin{abstract}
Conserved currents are discussed for static Conformal Killing Gravity, 
with explicit expressions in static spherical symmetry with anisotropic matter fluid or coupled to 
(non)linear electromagnetism. They are found in the reformulation 
of the third order equations by Harada as Einstein equations with
sources supplemented by a divergence-free anisotropic conformal
Killing tensor. A conserved current proposed by Altas and Tekin is also evaluated, and found nonzero for the vacuum solution by Harada.
\end{abstract}

\date{\today}

\maketitle

\section{Introduction}
A covariant definition of a conserved charge from a covariantly conserved current in a 
general spacetime was introduced by Aoki, Onogi and Yokoyama \cite{Aoki21}. A stress-energy
tensor with\protect{\footnote{Latin and Greek indices represent spacetime and space components.}} $\nabla_k T^k{}_j=0$ and a Killing vector,  
$\nabla_i\xi_j+\nabla_j \xi_i=0$,  provide a covariantly conserved current $J_k=\xi^l T_{kl}$ with the conserved charge
\begin{align}
Q=\int_{\Sigma_t} d^3 x \, \sqrt{-g(t,x)} J^0 (t,x) \label{QAoki}
\end{align}
$\Sigma_t$ is a time-slice of spacetime and $g$ is the determinant of the metric of the spacetime. 
For the Schwarzschild and the Reissner-Nordstr\"om black hole solutions, the 
singularity is treated as due to a delta-like source, so that \eqref{QAoki} provides the mass and the electric charge.

In this note we obtain conserved currents and charges in the frame of static Conformal Killing gravity (CKG), with explicit expressions in
spherical symmetry.\\
CKG was  introduced in 2023
by J. Harada \cite{Harada23,Harada23b}. It is characterized by the
field equations
\begin{align}
H_{jkl}=&8\pi G\,T_{jkl}\label{eq:Harada first}\\
H_{jkl}=&\nabla_jR_{kl}+\nabla_{k}R_{lj}+\nabla_{l}R_{jk} \nonumber\\
&-\tfrac{1}{3}(g_{kl}\nabla_jR+g_{lj}\nabla_{k}R+g_{jk}\nabla_{l}R) \label{HHH}\\
T_{jkl}=&\nabla_jT_{kl}+\nabla_{k}T_{lj}+\nabla_{l}T_{jk} \nonumber\\
&-\tfrac{1}{6}(g_{kl}\nabla_jT+g_{lj}\nabla_{k}T+g_{jk}\nabla_{l}T)\nonumber
\end{align}
where $R_{jk}$ is the Ricci tensor with trace $R$, $T_{kl}$ is
the stress-energy tensor with trace $T$. 
The Bianchi identity 
implies $\nabla_jT^{j}{}_{k}=0$. Solutions of the Einstein equations
are solutions of the new theory. 

Soon after, we found a
parametrization showing their equivalence to
the Einstein equations modified by a supplemental conformal Killing
tensor that is also divergence-free \cite{Mantica 23 a-1}:  
\begin{align}
 & R_{kl}-\tfrac{1}{2}Rg_{kl}=T_{kl}+K_{kl}\label{eq:einstein enlarged}\\
 & \nabla_jK_{kl}+\nabla_{k}K_{jl}+\nabla_{l}K_{jk} \nonumber\\
 &\qquad =\tfrac{1}{6}(g_{kl}\nabla_jK+g_{jl}\nabla_{k}K+g_{jk}\nabla_{l}K)\label{eq:Conformal Killing mantica-1}
\end{align}
For this reason the theory was given the current name.
The reformulation makes the extension of General Relativity (GR) explicit through the new term, that satisfies $\nabla^{k}K_{kl}=0$. As a source term, it may describe the still mysterious dark sector of gravity. \\
Inasmuch as a
Killing vector is a feature of a metric, 
a divergence-free conformal Killing tensor is a geometric entity peculiar of the metric.
In this sense (\ref{eq:Conformal Killing mantica-1}) is not an equation of motion, but rather a constraint.
%

For the Robertson-Walker metric
we found that $K_{ij}$ necessarily has the perfect fluid form, containing the scale factor but not its
derivatives \cite{Mantica 23 a-1}. Therefore the cosmological CKG theory is second order. 
In \cite{Mantica 24} we deepened its geometric
aspects and cosmological consequences. 

The majority of papers on CKG rely on the third-order Harada equations. They dealt with 
vacuum cosmological solutions \cite{Clement24}, spherically symmetric
solutions in vacuo \cite{Barnes23a}, with Maxwell source \cite{Barnes 24b}, vacuum Kundt solutions  \cite{Hervik 24},
black holes coupled to non-linear electrodynamics (NLE)
and scalar fields \cite{Junior24}, black-bounce solutions \cite{Junior 24 b}, wormholes with CKG black holes 
\cite{Alshal 24}.
Novel solutions were recently compared with the shadow of the supermassive black hole in the center of Milky Way \cite{Junior25}. \\
%
We obtained the parametrization of CKG in static spherically symmetric spacetimes
as an anisotropic conformal Killing tensor. This enabled us to reproduce in simple manner the known solutions 
for the vacuum and CKG coupled to NLE  \cite{Mantica 24b},  and study the junction conditions for
a spherical mass \cite{Mantica25}.

The issue of conserved charges in CKG in its third-order formulation \eqref{eq:Harada first}
has been scrutinized by Altas and Tekin \cite{Altas25}. As the field equations do not descend from an action principle, the theory is deprived of 
Noether's approach to symmetries \cite{Bajardi22} and the geometric tensor $H_{jkl}$ has zero divergence only when it solves the field equations. They object that it is an obstacle to interpretate integration constants, such as the mass parameter in the Schwarzschild metric, as conserved quantities. Energy and angular momentum of black hole solutions do not have a natural definition. For vacuum CKG with a Killing vector, the conserved current is identically zero in Einstein spacetimes, which are solutions of CKG.

In this note we study conserved currents in our parametrization \eqref{eq:Conformal Killing mantica-1} of static CKG spacetimes, where the field equations are second order and resemble those of GR. The currents are  proportional to the Killing vector of static spacetimes, and are nonzero also in absence of matter fields. 
Explicit expressions are given in spherical symmetry.

In Sect.~\ref{sec:Conserved-currents-derived} we recall some properties of the Killing vector and of conserved currents in static spacetimes. In Sect.~\ref{sec:Spherically-symmetric-static}, after a necessary recap of the conformal Killing and Ricci tensors  and the CKG equations in static spherically symmetric spacetimes,
we obtain the conserved currents arising from the matter stress-energy tensor and from the Killing tensor. 
In Sect.\ref{ATcurrent} we reconsider the correct current proposed by Altas and Tekin \cite{Altas25}, specify
it for static spacetimes and for the Harada vacuum. It is zero if the Killing tensor is zero. 
Conserved currents in CKG coupled to NLE are presented in Sect.~\ref{sec:NLE}. 

\section{\label{sec:Conserved-currents-derived}Conserved currents}
Given a symmetry with Killing vector $\nabla_i \xi_j +\nabla_j \xi_i=0$ and a conserved stress-energy tensor, $\nabla^{k}T_{kl}=0$, the vector field
\begin{equation}
J_l^{\sf mat} =T_{kl} \xi^k \label{eq:current from Killing}
\end{equation}
is a conserved current: $\nabla^l J^{\sf mat}_{l}=\xi^k\nabla^{l}T_{kl} +T_{kl}\nabla^{l}\xi^{k}=\frac{1}{2}T_{kl}(\nabla^{l}\xi^{k}+\nabla^{k}\xi^{l})=0$,  
as firstly reported in \cite{Trautman 58,Trautman 02,Fock 59,Komar 62}.
A conserved charge is then constructed, as shown for example in \cite{Feng 18}.\\ 
In the same way, in CKG a conserved current is 
\begin{align}
J_l^{\sf ckg}  = K_{kl} \xi^k  
\end{align}

In \cite{Komar 62,Komar 59} Arthur Komar introduced a conserved
current in generic spacetimes. For any vector $v^{j}$
the quantity (we follow the notation of J. Feng in \cite{Feng 18})
\begin{equation}
J^{\sf Kom}_j=\nabla_{l}(\nabla_j v^l - \nabla^l v_j )\label{eq:Komar current}
\end{equation}
is a covariantly conserved current: $\nabla_j J_{\sf Kom}^{j}$ $=\nabla_j\nabla_l (\nabla^jv^l - \nabla^l v^j) = (\nabla_j\nabla_l - \nabla_l\nabla_j )\nabla^j v^l = R_{jl}{}^j{}_m \nabla^m v^l$ $  -R_{lj}{}^{l}{}_m \nabla^j v^m = R_{lm}(\nabla^m v^l - \nabla^l v^m)=0$. Only the Ricci identity is used, the property does not depend upon the theory nor
the metric.\\ 
If $v^l$ is a Killing vector $\xi^l$, eq.(\ref{eq:Komar current}) takes the form 
$J^{\sf Kom}_j=-2\nabla_l \nabla^l \xi_j $.  
For Killing vectors it is known that 
%
$\nabla_{p}\nabla^{p}\xi_{l}=-R_{lm}\xi^{m}$ (see Appendix A). The Komar current becomes
\begin{equation}
J^l_{\sf Kom}=2R^l{}_m \xi^{m}\label{eq:Komar ricci}
\end{equation}
With the equation of motion \eqref{eq:einstein enlarged} the Komar current is related to the matter and CKG currents:
\begin{equation}
J_l^{\sf Kom}= R\xi_l +2 J_l^{\sf mat} + 2 J_l^{\sf ckg} 
\label{eq:Komar ricci stat}
\end{equation}
Indeed, one easily shows that also $R\xi_l$ is a conserved current. This fact, that the directional derivative of the curvature scalar along a Killing vector is zero, $\xi^k\nabla_k R=0$, was noticed by J. Feng (Appendix in \cite{Feng 18}).

\subsection*{Static spacetimes}
They are characterized by a timelike velocity $u_ju^j=-1$ and spacelike acceleration
$\dot u_k = u^j\nabla_j u_k$ such that (\cite{Stephani} p.283)
\begin{align}
\nabla_i u_j = -u_i \dot u_j , \qquad \nabla_i \dot u_j = \nabla_j \dot u_i \label{STATIC}
\end{align}  
A dot is a time derivative, $\dot \theta = u^k\nabla_k \theta$. We use the symbol $\eta=\dot u^k\dot u_k$.
The second equation implies $\dot u_j =\nabla_j \tau$, where $\tau $ is a scalar field with $\dot \tau=0$. \\
The vector $\xi_j = e^\tau u_j$ is a Killing vector for static spacetimes. In fact $\nabla_i \xi_j = \dot u_i \xi_j - u_j\xi_i$, so that $\nabla_i \xi_j +\nabla_j \xi_i =0$.  \\
In the static metric 
$$ds^2=-y(x)dt^{2}+g_{\mu\nu}^{\star}(x)dx^{\mu}dx^{\nu} $$
it is $u_0= -\sqrt y$, $u_\mu=0$,  $\sqrt y=e^\tau $ and $\xi^j=\delta^j{}_0$.\\
%
%
The property $R_{jk}u^k=-(\nabla^p\dot u_p) u_j$ is valid in all static spacetimes. It shows that the Killing vector is eigenvector of the Ricci tensor:
\begin{align}
R_{kl}\xi^l=-(\nabla_p\dot u^p)\xi_k \label{Riccixi}
\end{align}
Recall that in static spacetimes (eq.90 in \cite{Mantica 23 b})
\begin{align}
2\nabla_p\dot u^p = R^\star - R \label{RRstar}
\end{align}
where $R^\star$ is the scalar curvature of the space submanifold.  
If $G_{kl}=R_{kl}-\frac{1}{2} R g_{kl}$ is the Einstein tensor:
\begin{align}
G_{kl}\xi^l = -\frac{1}{2}R^\star \xi_k \label{GGGG}
\end{align}

For any scalar field $\theta $ with $\dot \theta= 0$ (such as $R$, $R^\star$, $\nabla_p\dot u^p$, $\eta$, ...) it is $\nabla_k (\theta \xi^k)=0$: such currents pertain to the geometry of static spacetimes. 
The contraction of the Killing vector with the zero-divergence stress-energy tensor is a conserved current of physical nature, that  
is linked to geometry by the field equations. In CKG the link is
\begin{align}
-\frac{1}{2}R^\star \xi_k = J_k^{\sf mat} + J_k^{\sf ckg}
\end{align}

In static spacetimes the natural vectors for a Komar current are $u^l$ and $\dot u^l$. The second one is closed and gives no Komar current. 
The first one gives $J^j_{\sf Kom} =\nabla_l(\nabla^j u^l - \nabla^l u^j) = \nabla_l (-u^j\dot u^l + u^l\dot u^j) = 
u_l \dot u^j\dot u^l - u_j \nabla_l\dot u^l + (\nabla_l u^l) \dot u^j + u^l\nabla_l \dot u^j$. Now use $u_l\dot u^l=0$,  $\nabla_l u^l=0$ and $u^l\nabla_l \dot u_j = u^l \nabla_j \dot u_l =  -\dot u_l \nabla_j u^l = u_j \dot u^l\dot u_l$. Then:
$$ J^{\sf Kom}_j = u_j (\eta - \nabla_l \dot u^l) $$ 
With the Killing vector $\xi_j = e^\tau u_j$, the Komar current is geometric, as above discussed.

\subsection*{\label{sec:StaticSpherical} Static spherical symmetry}
Consider the static metric with spherical symmetry
\begin{align}
ds^2=-y(r)dt^{2}+\dfrac{h^{2}(r)}{y(r)}dr^{2}+r^2 (d\theta^2 +\sin^2\theta \,d\phi^2) \label{eq:Static sph symm}
\end{align}
The metric tensor has the structure
\begin{align}
g_{kl}= -u_k u_l + \frac{\dot u_k\dot u_l}{\eta} + N_{kl} \label{gtensor}
\end{align}
where $N_{kl}$ is the projection on the submanifold orthogonal to $u_k$ and $\dot u_k$.\\
Besides the time-like Killing vector $\xi^k$, there are three Killing vectors $x_i\partial_j -x_j\partial_i$ 
associated to rotations. In the spherical coordinates ($t,r,\theta,\phi)$ their
covariant components are \cite{DEFELICE}:
\begin{align*}
& \zeta_k = r^2(\sin\phi \, \delta_{k,\theta} + \cos\theta\sin\theta \cos\phi \, \delta_{k,\phi})  \\
& \eta_k =r^2 (\cos\phi \,\delta_{k,\theta} - \cos\theta\sin\theta \sin\phi \, \delta_{k,\phi}) \\
& \varphi_k  = r^2 \sin^2\theta \, \delta_{k,\phi}
\end{align*}
They are orthogonal to $u_k$ and $\dot u_k$ and linearly dependent. 

The totally reducible Killing tensor 
\begin{align}
\Omega_{kl} = \zeta_k\zeta_l +\eta_k\eta_l+\varphi_k\varphi_l =
r^4(\delta^\theta_k \delta^\theta_l +\sin^2\theta\, \delta^\phi_k \delta^\phi_l) \label{OMEGA}
\end{align} 
is diagonal and
proportional to the angular part of the metric tensor: 
$\Omega_{kl} =  r^2 N_{kl} $.

\section{\label{sec:Spherically-symmetric-static}Static spherically symmetric CKG}
In static spherically symmetric spacetimes, the covariant expression of the Ricci tensor is eq.85 in \cite{Mantica 23 b}:
\begin{align}
R_{kl}=\frac{R+4\nabla_{p}\dot u^{p}}{3}u_{k}u_l+\frac{R+\nabla_{p}\dot u^{p}}{3}g_{kl} \nonumber\\
+\Sigma(r)\left[\frac{\dot u_{k}\dot u_l}{\eta}-\frac{u_{k}u_l+g_{kl}}{3}\right]. \label{eq:Ricci static}
\end{align}
Given the expression of the metric \eqref{eq:Static sph symm}, 
one evaluates $\nabla_p \dot u^p$, $\Sigma (r)$ and $R$ related to $R^\star$ by \eqref{RRstar}:
\begin{align}
 & \nabla_p \dot u^p = \frac{y''}{2h^2}-y'\frac{h'}{2h^3}+\frac{y'}{rh^2} \label{eq:diva acc}\\
 & \Sigma(r)=-\frac{y''}{2h^2}+y'\frac{h'}{2h^3}+\frac{1}{r^2}\frac{y}{h^2}-\frac{1}{r^2}+\frac{y}{r} 
 \frac{h'}{h^3} \label{eq:anisotropi factor}\\
& R^{\star}=\frac{2}{r^2}+\frac{4y}{r} \frac{h'}{h^3}-\frac{2y'}{rh^2}-\frac{2}{r^2}\frac{y}{h^2}\label{eq:Scalar curvatures}
\end{align}
They are eqs. 87, 89 and 91 in \cite{Mantica 23 b} with $y=b^2$; a prime is a derivative in $r$.
In  \cite{Mantica 24b} we proved the following 
\begin{prop}
\label{prop 5 Anisotropic} In the spherically symmetric static spacetime
\eqref{eq:Static sph symm}, the tensor
\begin{equation}
K_{kl}={\sf A}(r)u_{k}u_l+{\sf B}(r)g_{kl}+{\sf C}(r)\frac{\dot u_{k}\dot u_l}{\eta} \label{eq:anisotropic Conformal killing}
\end{equation}
is divergence-free and conformal Killing if and only if 
${\sf A}(r)=\kappa_2 r^2-2\kappa_3 y(r)$, \\
${\sf B}(r)=\kappa_1+2\kappa_{2}r^{2}+\kappa_{3}y(r)$,\\
${\sf C}(r)=-\kappa_2 r^2$,\\
%
where $\kappa_{1,}\kappa_{2,}\kappa_{3}$ are constants. 
\end{prop}
\noindent
Note the absence of $h$ in the formulas. Since $K_{ij}$ does not contain derivatives of the metric function $y$, the theory is second order in derivatives of the metric.

\begin{remark}
As noted by Barnes \cite{BarnesPP}, the conformal Killing tensor yields a Killing tensor $K^{(0)}_{ij} =K_{ij} - \frac{1}{6}Kg_{ij}$. 
With eqs.\eqref{gtensor} and \eqref{OMEGA} it is a combination of Killing tensors:
\begin{align}
K^{(0)}_{ij} = -2\kappa_3 \xi_i\xi_j +\frac{1}{3}\kappa_1 g_{ij} + \kappa_2 \Omega_{ij}
\end{align}
\end{remark}

%
Given the tensor form of $R_{ij}$ and $K_{ij}$, the CKG field equations $G_{kl}= T_{kl}+K_{kl}$ determine the stress energy tensor
in the form of an anisotropic fluid 
\begin{align}
T_{kl}=& R_{kl} - \frac{R}{2}g_{kl} - {\sf A}u_{k}u_l-{\sf B}g_{kl}-{\sf C}\frac{\dot u_k\dot u_l}{\eta}\\
=&(\mu+p_{\perp})u_{k}u_l+p_{\perp}g_{kl}+(p_{r}-p_{\perp})\frac{\dot u_k\dot u_l}{\eta}\nonumber
\end{align}
where $\mu$ is the matter energy density, $p_{r}$ and $p_{\perp}$ are
the radial and transverse pressures. 
Equality of the coefficients of the basic tensors $u_i u_j$, $\dot u_i\dot u_j$ and $g_{ij}$ yield three scalar field equations for CKG (\cite{Mantica 24b} eq. 15), which combine to give
\begin{align}
&\frac{R^\star}{2}=\mu-\kappa_1 -\kappa_2 r^2 -3\kappa_3 y(r)\\
&\nabla_p \dot u^p=\frac{1}{2}(p_r+2p_\perp)+\frac{\mu}{2}+\kappa_1+2\kappa_2 r^2 \\
&\Sigma=(p_r - p_\perp)-\kappa_2 r^2
\label{eq: field with functions}
\end{align}

With a divergence-free energy momentum tensor for the matter fluid, and a divergence-free Killing tensor,
there are two conserved currents associated to the Killing vector $\xi_j=e^\tau u_j$.\\
The first one is $J^{\sf mat}_l=T_{jl} \xi^j =-\xi_l \mu $, with conserved charge \eqref{QAoki} that is the energy of the matter fluid. Recalling that
$\xi^j =\delta^j{}_0$ in the coordinates \eqref{eq:Static sph symm}:
\begin{align}
E_{\sf mat}
= 4\pi \int r^2 h(r)\mu (r) \, dr
\end{align}
The other conserved current stems from the divergence-free Killing tensor: 
\begin{align}
J^k_{\sf ckg} =K^k{}_j \xi^j  =  \xi^k (\kappa_1+\kappa_2 r^2 +3 \kappa_3 y(r)) \label{darkJ}
\end{align}
with conserved `dark energy' 
\begin{equation}
E_{\sf ckg}  = 4\pi \int r^2 h(r)  [\kappa_1+\kappa_2 r^2+3\kappa_3 y(r)] dr
\end{equation}

A specific evaluation is now done for the Harada's vacuum solution ($T_{kl}=0$): 
\begin{equation}
y (r) =1-\dfrac{2M}{r}-\frac{\Lambda}{3}r^{2}-\frac{\lambda}{5} r^4 \label{eq:Harada vacuum}
\end{equation}
As shown in \cite{Mantica 24b} it arises when $h(r)=1$, $\kappa_1=-\Lambda$, $\kappa_2=-\lambda$ 
and $\kappa_3=0$.\\ 
With zero Killing tensor it is the Schwarzschild metric $y =1- 2M/r$ of GR. They are both singular in $r=0$. Aoki et al.  have shown in \cite{Aoki21} eq.12, that, despite $T_{kl}$ being zero, the parameter $M$ 
is interpretable in the distributional sense as a conserved mass-charge (the total energy).
We apply the same reasoning to the Harada solution of CKG. Consider the conserved current \eqref{GGGG}: $J_k = G_{jk}\xi^j =-\frac{1}{2}R^\star \xi_k$. With eq.\eqref{eq:Scalar curvatures}:
$$J_k = -\frac{1}{2} R^\star \xi_k = -\left[\frac{1}{r^2} - \frac{y'}{r} -\frac{y}{r^2} \right ] \xi_k $$ 
The metric function \eqref{eq:Harada vacuum} is sum of a regular term $y_{reg}=1-(\Lambda/3)r^2 - (\lambda/5)r^4$ 
and of a singular one $y_{sing}=-2M/r$. Then:
\begin{align*}
J^k =& \left[ \frac{y'_{sing}}{r} +\frac{y_{sing}}{r^2} \right ] \xi^k 
+ \left[ \frac{y'_{reg}}{r} + \frac{y_{reg}-1}{r^2} \right ] \xi^k \\
=& \frac{1}{r^2} \frac{d}{dr}(r y_{sing})\xi^k 
- ( \Lambda  + \lambda r^2  )\xi^k \\
=& J^k_{sing} + J^k_{\sf{ckg}}
\end{align*} 
The second term is conserved, it is \eqref{darkJ}. With $y_{sing}=-2M/r$ the first one would be zero. However, a point 
mass is conceivable as a distribution. We enforce a step function $\theta (r)$ in the solution to gain the sensible result
\begin{align}
 J^k_{sing} = \xi^k \frac{1}{r^2} \frac{d}{dr}(-2M\theta (r) ) = -\frac{2M}{r^2} \delta(r) \xi^k 
 \end{align}
with associated conserved charge 
\begin{align}
E=4\pi \int dr r^2 J_{sing}^0 = 8\pi M
\end{align}
The energy cost for a Harada black hole is identical to the Schwarzschild case.\\
Another charge comes from the Killing sector:
\begin{equation}
E_{\sf Harada}= -4\pi \int_{0}^{\bar r}(\lambda r^{4}+\Lambda r^{2})dr\label{eq:cons charge harada' vacuum}
\end{equation}
If $\lambda>0$ in \eqref{eq:Harada vacuum}, positivity of $y$ requires the presence of an outer horizon at $\bar r$, that guarantees the boundedness of the energy.

\section{A current by Altas \& Tekin\\
(from the III order field equations)}\label{ATcurrent}
In \cite{Altas25} Altas and Tekin evaluated the double divergence of the tensor $H_{jkl}$ in \eqref{HHH}
and wrote it as a divergence of a symmetric tensor: $ \nabla^k \nabla^j H_{jkl}=\nabla^{k}\Phi_{kl} $. 
We redo and correct their calculation. 
The starting point is eq.41 in \cite{Altas25}:
\begin{align*}
\nabla^{k}\nabla^jH_{jkl}&=6R_{km}\nabla^k R_l{}^m - 3R^{mj} \nabla_lR_{mj}\\
&+\tfrac{1}{2}\nabla_{l}(\nabla^{2}R)+\tfrac{7}{3} R_l{}^m \nabla_m R +4R_{jklm}\nabla^k R^{mj} \nonumber
\end{align*}
Substitution of the Einstein tensor $G_{jm}=R_{jm}-\frac{1}{2}Rg_{jm}$
(we drop the cosmological constant present in \cite{Altas25}, since the conformal Killing tensor already
contains it) gives 
\begin{align}
&\nabla^k \nabla^j H_{jkl}= 6 \, G^{km} \nabla_k G_{lm} - 3 G^{mj} \nabla_l G_{mj}
+\tfrac{1}{2} \nabla_l (\nabla^2 R)\nonumber \\  
&+\tfrac{10}{3}G_{kl} \nabla^{k}R + \tfrac{5}{3}R\nabla_l R 
+4R_{jklm}\nabla^k G^{mj}\label{eq:Bidiv with G}
\end{align}
This equation corrects eq.44 in \cite{Altas25}.\\
We continue their procedure and rewrite it:
\begin{align}
&\nabla^{k}\nabla^jH_{jkl}=\nabla^{k}( 6\, G_{km}G_{l}^{m}+\tfrac{10}{3}RG_{kl}) \label{eq:Bidiv with G 2}\\
&+\nabla_l(-\tfrac{3}{2}G_{mj}G^{mj} + \tfrac{1}{2}\nabla^2 R +\tfrac{5}{6}R^2 ) 
+4R_{jklm}\nabla^{k}G^{mj} \nonumber
\end{align}
The equation contains a divergence and a gradient. The last term
may be decomposed in the same manner: 
\begin{align*}
&R_{jklm} \nabla^k G^{mj}  
=\nabla^{k}(R_{jklm}G^{mj})-G^{mj} \nabla^{k}R_{lmjk} \\
&=\nabla^{k}(R_{jklm}G^{mj})-G^{mj}(\nabla_mR_{kj}-\nabla_{l}R_{mj})\\
&=\nabla^{k}(R_{jklm}G^{mj})-G^{mj}(\nabla_mG_{lj}+\tfrac{1}{2} g_{lj} \nabla_m R )\\
&\quad +G^{mj}(\nabla_l G_{mj}+\tfrac{1}{2} g_{mj}\nabla_l R)\\
&=\nabla^{k}(R_{jklm}G^{mj})- G^{mj} \nabla_mG_{lj}-\tfrac{1}{2}\nabla_m(RG_{l}^{m})\\
&\quad +G^{mj}\nabla_l G_{mj} -\tfrac{1}{2}R \nabla_l R\\
&=\nabla^{k}(R_{jklm}G^{mj})-\nabla_m(G_{lj}G^{mj})\\
&\qquad -\tfrac{1}{2}\nabla_m(RG_{l}^{m})+\tfrac{1}{2}\nabla_{l}(G_{mj}G^{mj})-\tfrac{1}{4}\nabla_{l}R^{2}
\end{align*}
Eq.\eqref{eq:Bidiv with G 2} gains the form of a total divergence 
\begin{align}
\nabla^k \nabla^j H_{jkl} = &\nabla^k \left ( 4R_{jklm}G^{mj}+2G_{km}G_l{}^m +\tfrac{4}{3} R G_{kl} \right )\nonumber \\
&+\nabla_l \left ( -\tfrac{1}{6} R^2 +\tfrac{1}{2} \nabla^2 R+ \tfrac{1}{2} G_{pq}G^{pq} \right ) \label{eq:Bidiv final}
\end{align}
This replaces eq.49 in \cite{Altas25}. We read the tensor $\Phi_{kl}$, that replaces the expression eq.50 in \cite{Altas25}:
\begin{align}
\Phi_{kl}=&g_{kl}\left (\tfrac{1}{2}G_{pq}G^{pq}-\tfrac{1}{6}R^{2}+\tfrac{1}{2}\nabla^{2}R\right ) \nonumber \\
&+4R_{jklm}G^{mj}+2G_{km}G_{l}^{m}+\tfrac{4}{3}RG_{kl}\label{eq:conserved tensor}
\end{align}

Vacuum solutions of the field equations \eqref{eq:Harada first} solve $H_{jkl}=0$, then
$\nabla^k\Phi_{kl}=0$, and a conserved current $\Phi_{kj}\xi^j$ is found.\\
In Einstein spacetimes $R_{jm}=\frac{1}{4}Rg_{jm}$. Then
$2 G_{km}G^m{}_l$ $=\frac{1}{8}R^2g_{kl}$, $4R_{jklm}G^{mj}=
\frac{1}{4}R^2g_{kl}$.
Thus $\Phi_{kl}=0$ identically, and the associated conserved current
vanishes. This fact does not occur for eq.50 in \cite{Altas25}.\\

Now we continue, and evaluate the Altas-Tekin conserved current in static spacetimes. Using eq.\eqref{GGGG}: 
\begin{align}
\Phi_{kl}\xi^l=&\xi_k \left[\tfrac{1}{2}G_{pq}G^{pq}-\tfrac{1}{6}R^{2}+\tfrac{1}{2}\nabla^{2}R +\tfrac{1}{2} R^{\star 2}
 -\tfrac{2}{3}RR^\star  \right] \nonumber \\
&+4R_{jklm}G^{mj}\xi^l 
\end{align}
The last term is evaluated with eq.\eqref{STATIC}:
\begin{align*}
&G^{jm}R_{jklm}\xi^l = -e^\tau G^{jm}(\nabla_j\nabla_k-\nabla_k \nabla_j)u_m \\
&= e^\tau  G^{jm}[\nabla_j (u_k \dot u_m) - \nabla_k (u_j \dot u_m)]\\
&= - G^{jm}\xi_j (\dot u_k \dot u_m +\nabla_k \dot u_m) + \xi_k G^{jm} (\dot u_j\dot u_m  +  \nabla_j \dot u_m)\\
& = \tfrac{1}{2}R^\star \xi^m (\dot u_k \dot u_m +\nabla_m \dot u_k) 
+ \xi_k G^{jm} (\dot u_j\dot u_m  +  \nabla_j \dot u_m) \\ 
&= \tfrac{1}{2}R^\star e^\tau  \ddot u_k + \xi_k G^{jm} (\dot u_j\dot u_m  +  \nabla_j \dot u_m) \\
&= \xi_k [\tfrac{1}{2}R^\star \eta +  G^{jm} (\dot u_j\dot u_m  +  \nabla_j \dot u_m) ]
 \end{align*}
 where we used $\ddot u_k =\eta u_k$.\\
 We obtain a current proportional to the Killing vector:
 \begin{align}
&\Phi_{kl}\xi^l= \varphi \xi_k\\
&\varphi=\tfrac{1}{2}G_{pq}G^{pq}-\tfrac{1}{6}R^{2}+\tfrac{1}{2}\nabla^{2}R +\tfrac{1}{2} R^{\star 2} \nonumber\\
&\quad -\tfrac{2}{3}RR^\star +2R^\star \eta +  4G^{jm} (\dot u_j\dot u_m  +  \nabla_j \dot u_m) \nonumber
 \end{align}
 
 In CKG vacuum it is $G_{jk}=K_{jk}$. With the Killing tensor 
 eq.\eqref{eq:anisotropic Conformal killing}:
 \begin{align*}
\varphi &= \tfrac{1}{2}[{\sf A}^2 - 2{\sf AB} +4{\sf B}^2 + 2{\sf BC} +{\sf C}^2]\\
&-\tfrac{1}{6}R^{2}+\tfrac{1}{2}\nabla^{2}R +\tfrac{1}{2} R^{\star 2} -\tfrac{2}{3}RR^\star \\
& +  2(2{\sf B} + 2{\sf C}+R^\star -2{\sf A})\eta + 4{\sf B}\nabla_p\dot u^p + 2{\sf C} \dot u^j \nabla_j \log\eta 
 \end{align*}
For the Harada vacuum eq.\eqref{eq:Harada vacuum}: ${\sf A}= -\lambda r^2$, ${\sf B}=-\Lambda -2\lambda r^2$, 
${\sf C}=\lambda r^2$, $R=4\Lambda +6\lambda r^2$, $R^\star = 2\Lambda +2\lambda r^2$, $\nabla_p\dot u^p =
{\sf B}$. One obtains:
  \begin{align*}
\varphi (r)=& [5\lambda^2r^4 + 6\Lambda \lambda r^2 +2 \Lambda^2]  -\tfrac{2}{3}(2\Lambda +3\lambda r^2)^2\\
&+3\lambda\nabla^2 r^2 +2(\Lambda +\lambda r^2)^2 -\tfrac{8}{3} (2\Lambda +3\lambda r^2)(\Lambda +\lambda r^2) \\
&+4\lambda r^2\eta  +4(\Lambda +2\lambda r^2)^2 + 2 \lambda r^2  \dot u^j \nabla_j \log\eta
\end{align*}
$\nabla^2 r^2$ is evaluated with the Laplace-Beltrami operator ($h(r)=1$: $|g|=r^4\sin^2\theta $, $g^{rr}=y(r)$): 
\begin{align*}
\frac{1}{\sqrt {|g|}} \frac{d}{dr}\left[ \sqrt{|g|} g^{rr} \frac{d}{dr} r^2\right ] =&  \frac{2}{r^2} \frac{d}{dr}(r^3 y) =
6y +2ry'
\end{align*}
We also need the term
\begin{align*}
g^{rr}\dot u_r \frac{d}{dr} \log\eta = y \frac{y'}{2y} \frac{d}{dr} \log \frac{y'^2}{4y}  = y'' - 2\eta 
\end{align*}
The conclusion is:
\begin{align}
\varphi (r) =
 -\frac{32M\lambda}{r} +18 \lambda -\frac{20}{3}\Lambda\lambda r^2 - \frac{21}{5} \lambda^2 r^4 .
\end{align}
The current is nonzero. It vanishes for $\lambda =0$, while the Killing tensor reduces to a cosmological term 
$-\Lambda g_{jl}$, and the Harada vacuum solution becomes Schwarzschild - de Sitter.
 
\section{(Non)Linear electrodynamics}\label{sec:NLE}
Consider CKG coupled to static non-linear electrodynamics (see for
example \cite{Mantica 24b,Junior24}). The stress energy tensor is 
\begin{align}
T_{kl}=2(\mathbb{E}^{2}+\mathbb{B}^{2})\mathcal{L}_{F}(F)\left[u_{k}u_l-\frac{\dot u_{k}\dot u_l}{\eta}\right] \nonumber\\+2\left[\mathbb{B}^{2}\mathcal{L}_{F}(F)-\mathcal{L}(F)\right] g_{kl}
\end{align}
where $F=\frac{1}{4} F_{pq} F^{pq} = \frac{1}{2} (\mathbb{B}^2 - \mathbb{E}^2)$
is the Faraday scalar, $\mathbb{E}$ and $\mathbb{B}$ are the electric and magnetic fields, $\mathcal{L}(F)$ is the Lagrangian density, and $\mathcal{L}_F= d\mathcal{L} /dF$. It is known that
$$\mathbb{E}=\dfrac{q_{e}}{r^{2}\mathcal{L}_{F}(F)}, \qquad \mathbb{B}=\dfrac{q_{m}}{r^{2}}$$
with charges $q_e$ and $q_m$.  CKG solutions are displayed in \cite{Barnes 24b,Clement24,Mantica 24b}. 
A conserved current is 
\begin{equation}
J^{\sf nLed}_k= T_{kl}\xi^l = [-2\mathbb{E}^2 \mathcal{L}_F (F)-2 \mathcal{L}(F)] \xi_k \label{eq:cons current non  lin elect}
\end{equation}
The Killing tensor mantains the form \eqref{eq:anisotropic Conformal killing}, so that another conserved current  is
$J_k^{\sf ckg}$ in eq.\eqref{darkJ}.

In linear electrodynamics $\mathcal{L}(F)=F=\frac{1}{2}(\mathbb{B}^{2}-\mathbb{E}^{2})$, so that $\mathbb{E}=q_e /r^2$.
The conserved current density is
\begin{equation}
J^{\sf Led}_k = -\frac{q_e^2+q_m^2}{r^4} \xi_k \label{eq:conserved current  linear electrodyn}
\end{equation}
%
%
%

%
Let us consider NLE with $\mathbb{E}=0$. 
In \cite{Junior24} the non-linear Lagrangian density is reconstructed
given an input metric as in the following examples with $h=1$. These are two examples, with the metric and the 
generating Lagrangian.\\
Example 1: The Bardeen-type metric
\begin{gather}
y=1-\dfrac{2Mr^{2}}{(q_{m}^{2}+r^{2})^{3/2}}-\dfrac{\Lambda}{3}r^{2}-\dfrac{\lambda}{5}r^{4}, \\ 
\mathcal{L}=\dfrac{3Mq_{m}^{2}}{(q_{m}^{2}+r^{2})^{5/2}}+\phi_{1}r^{2}+\phi_{0}
\end{gather}
Example 2: The Hayward-type metric (firstly proposed
in \cite{Mantica 24b}) 
\begin{gather}
y=1-\dfrac{2Mr^{2}}{q_{m}^{3}+r^{3}}-\dfrac{\Lambda}{3}r^{2}-\dfrac{\lambda}{5}r^{4}, \\
\mathcal{L}=\dfrac{3Mq_{m}^{2}}{(q_{m}^{3}+r^{3})^{2}}+\phi_{1}r^{2}+\phi_{0}
\end{gather}

\noindent
$\phi_1$ and $\phi_0$ are constants.
The conserved current densities can be explicitly evaluated with eq.\eqref{eq:cons current non  lin elect}, that simplifies to
$J^{\sf nLed}_k=-2\mathcal{L}\xi_k$.

\section{Conclusions }
The 2nd order parametrization of the 3rd order Harada equations in static spacetimes enables a direct identification of
conserved currents. They arise as the standard contraction of the divergence-free stress energy tensor and of the conformal Killing tensor, with the time-like Killing vector. 
The mass parameter of Harada's vacuum solution is clarified in the distributional sense, as shown by Aoki et al. \cite{Aoki21}. 
We correct and further implement a conserved current by Altas and Tekin, derived in the third order formalism. We find that, contrary to their claim, it is in general nonzero.  \\
In  CKG coupled to (non)LE, the current related to the Faraday tensor is the same as in GR coupled to (non)LE.

\section*{Appendix A}
%
If $\xi_{j}$ is a Killing vector, then 
$$\nabla_j\nabla_{k}\xi_{l}=-R_{klj}\,^{m}\xi_{m}$$
{\em Proof}:
A covariant derivative of the Killing condition $\nabla_{k}\xi_{l}+\nabla_{l}\xi_{k}=0$
reads $\nabla_j\nabla_{k}\xi_{l}+\nabla_j\nabla_{l}\xi_{k}=0$.
The Ricci identity allows to rewrite it as 
\begin{align*}
\nabla_j\nabla_{k}\xi_{l}+\nabla_{l}\nabla_j\xi_{k}+R_{jlk}\,^{m}\xi_{m}=0
\end{align*}
A cyclic permutation of the indices brings the equations 
\begin{align*}
\nabla_{k}\nabla_{l}\xi_{j}+\nabla_j\nabla_{k}\xi_{l}+R_{kjl}\,^{m}\xi_{m}=0\\
\nabla_{l}\nabla_j\xi_{k}+\nabla_{k}\nabla_{l}\xi_{j}+R_{lkj}\,^{m}\xi_{m}=0
\end{align*}
Now add the first and second equations and subtract the third one:
\begin{align*}
2\nabla_j\nabla_{k}\xi_{l}+(R_{jlk}\,^{m}+R_{kjl}\,^{m}-R_{lkj}\,^{m})\xi_{m}=0
\end{align*}
The first Bianchi identity gives the result.
A contraction with the metric tensor gives $\nabla_{p}\nabla^{p}\xi_{l}=-R_{lm}\xi^{m}$. \hfill $\square $
\end{document}